\documentclass[11pt,a4paper]{article}
\usepackage{hlc}
\usepackage{tcolorbox,listings}
\usepackage{colortbl}
\usepackage{tabularx}

\tcbset{geotask/.style={}}

\definecolor{geoLayerBg}{HTML}{DAE8FC}
\definecolor{geoActionBg}{HTML}{FFE6CC}
\definecolor{geoSignalBg}{HTML}{D5E8D4}
\definecolor{geoHeaderBg}{HTML}{EEF1F6}
\definecolor{geoInk}{HTML}{26354D}
\definecolor{geoRule}{HTML}{9DAEC3}

\tcbset{geotask/.style={float=tb}}
\colorlet{geoLayerBg}{hlcBlueSoft}
\colorlet{geoActionBg}{hlcAmberSoft}
\colorlet{geoSignalBg}{hlcTealSoft}
\colorlet{geoHeaderBg}{hlcPaper}
\colorlet{geoInk}{hlcInk}
\colorlet{geoRule}{hlcLine}
\colorlet{madder}{hlcSelect}
\newcolumntype{L}[1]{>{\raggedright\arraybackslash}p{#1}}

\BeforeBeginEnvironment{lstlisting}{\par\noindent\minipage{\linewidth}}
\AfterEndEnvironment{lstlisting}{\endminipage\par}
\let\hlcoriginalincludegraphics\includegraphics
\RenewDocumentCommand{\includegraphics}{O{} m}{%
  \IfFileExists{hlc/#2}{\hlcoriginalincludegraphics[#1]{hlc/#2}}%
    {\hlcoriginalincludegraphics[#1]{#2}}}
\title{Humanity's Last Choice:\\
A Common Evaluation Standard for Generative Engine Optimization}

\hlcauthor[1,*]{Qianfeng Wen}
\hlcauthor[1,*]{Yifan Simon Liu}
\hlcauthor[1]{Difan Jiao}
\hlcauthor[2]{Junda Wu}
\hlcauthorbreak
\hlcauthor[3]{Di Bai}
\hlcauthor[4]{Jintao Liu}
\hlcauthor[4]{Xidong Wu}
\hlcauthor[1,4]{Zhenwei Tang}
\hlcaffiliation{1}{University of Toronto}
\hlcaffiliation{2}{University of California, San Diego}
\hlcaffiliation{3}{Google DeepMind}
\hlcaffiliation{4}{Google}
\hlccontribution{\hlcmark{*}Equal contribution.}
\hlclinks{%
  \hlclink{\faGithub}{github.com/humanitylastchoice}{https://github.com/humanitylastchoice}%
  \hlclink{\faGlobe}{humanitylastchoice.org}{https://humanitylastchoice.org}%
  \hlclink{\faEnvelope[regular]}{info@humanitylastchoice.org}{mailto:info@humanitylastchoice.org}}
\newcommand{\teaserref}{}
\begin{document}
\maketitle
\begin{abstract}
\emph{Humanity's Last Choice} describes the idea that the last choice people need to make is which AI system they entrust with the decisions that follow. These systems rely on external sources that content providers can edit through generative engine optimization (GEO) to increase source visibility and influence recommendations. Choosing among these systems therefore requires comparable evidence of how well they carry
out user requests when exposed to GEO. Benchmarks have made progress in standardizing tasks, models, and metrics, but they often define their own spaces of GEO strategies and use different editing constraints. These differences can conflate gains from better optimization algorithms with those from greater freedom to edit sources, while making evaluation coverage difficult to compare across benchmarks.
We therefore call for a common evaluation standard and propose a meta-taxonomy that describes where sources are edited, how they are edited, and what signals those edits target. Drawing on representative literature, we map existing strategies into a searchable space and outline a protocol for controlled comparisons with explicit coverage reporting. This provides a foundation for common leaderboards that can inform which systems people choose to rely on.
\end{abstract}


\begin{figure*}[t]
\centering
\includegraphics[width=\textwidth]{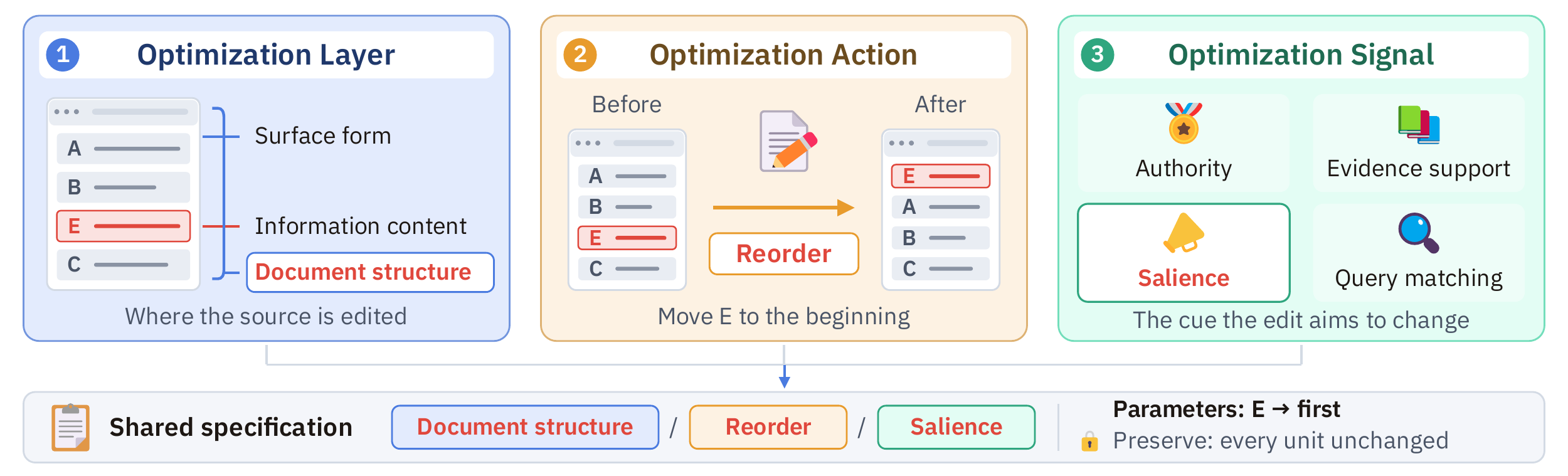}
\caption{From a GEO strategy to a shared specification. Moving the existing evidence unit $E$ to the beginning changes document structure (\emph{Optimization Layer}) through reordering (\emph{Optimization Action}) and targets salience (\emph{Optimization Signal}). The specification identifies the unit and its destination and requires all units, including $A$, $B$, and $C$, to remain intact. The displayed layers and signals illustrate part of the taxonomy.}
\label{fig:overview}
\end{figure*}

\section{Introduction}
\label{sec:intro}

Conversational agents such as ChatGPT and personal agents such as
OpenAI's dots, Meta's Muse, and Google's Gemini Spark are increasingly
capable of gathering information, comparing options, and carrying out
tasks on a user's behalf
\citep{openai2026dots,meta2026muse,google2026spark}.
As their capabilities grow, people can delegate more of the decisions
involved in those tasks to them.
\emph{Humanity's Last Choice} describes the idea that the last choice
people need to make is which AI system to entrust with the decisions
that follow.\footnote{The name echoes Humanity's Last Exam
\citep{hle2026nature}, which likewise evaluates AI on a problem
vital to humanity, but through expert academic questions.} Such a system evaluates information and recommends
candidates using sources that content providers can edit.

Generative engine optimization (GEO) studies how such edits
affect source visibility in generated answers
\citep{aggarwal2024geo,kumar2024manipulating,pfrommer2024ranking}.
Choosing which system to rely on therefore requires comparable
evidence of how well it carries out user requests when those sources
are optimized to influence its judgments\teaserref.
GEO strategies have grown beyond a fixed list of writing tips. They include authoritative phrasing, added
citations, appended token sequences, and document-level feature
changes. AutoGEO learns engine preference rules, while AgenticGEO
evolves a reusable strategy archive
\citep{wu2025autogeo,yuan2026agenticgeo}.

Existing benchmarks, including C-SEO Bench, E-GEO, and the
ranking-manipulation GEO-Bench, have made progress in standardizing
tasks, models, and metrics
\citep{puerto2025cseo,bagga2025egeo,nimase2026geobench}.
However, shared tasks, models, and metrics do not by themselves
define the strategies that methods are allowed to apply.
A method limited to appending a short sequence and another allowed
to rewrite an entire page can be evaluated by the same ranker but
operate under different editing permissions. Comparing their scores
without accounting for these differences can conflate gains from
better optimization algorithms with those from greater freedom
to edit sources. These differences also affect system selection
because stronger task performance may reflect a less demanding set
of edits. A common evaluation standard therefore needs to specify
both the strategy space being evaluated and the conditions under
which methods explore it.

Our position is that \textbf{a common evaluation standard for GEO
should be grounded in a shared specification of where sources are
edited, how they are edited, and what signals the edits
target.} We propose a meta-taxonomy with three dimensions.
\emph{Optimization Layer} identifies the source layer being modified,
such as wording, information content, or document structure.
\emph{Optimization Action} specifies the operation applied, such as
inserting, deleting, or reordering.
\emph{Optimization Signal} identifies the cue through which the edit aims to influence the model's judgment,
such as authority or salience
\mbox{(Figure~\ref{fig:overview})}.

Drawing on representative methods, we map their
strategies into the space.
We discuss how the space can support strategy construction, search,
and evaluation under shared settings. These uses motivate a
community agenda for extending the space as new strategies
emerge and for developing GEO leaderboards across methods, models,
and domains. Such comparisons can provide a clearer basis for
choosing AI systems for the tasks people intend to delegate to them.

\begin{table*}[t]
\centering
\begingroup
\fontsize{9}{10.3}\selectfont
\setlength{\tabcolsep}{4pt}
\setlength{\extrarowheight}{0.4pt}
\renewcommand{\arraystretch}{1.0}
\arrayrulecolor{geoRule}
\newcommand{\geoTableRef}[2]{\mbox{#1~\citeyearpar{#2}}}
\begin{tabularx}{\textwidth}{@{}l
  >{\raggedright\arraybackslash}X
  >{\raggedright\arraybackslash}p{0.29\textwidth}@{}}
\toprule
\rowcolor{geoHeaderBg}
 & \textcolor{geoInk}{\textbf{Description}} &
 \textcolor{geoInk}{\textbf{Literature}} \\
\midrule
\rowcolor{geoLayerBg}
\multicolumn{3}{@{}p{\textwidth}@{}}{\rule{0pt}{11.5pt}\textcolor{geoInk}{\textbf{Optimization Layer} \hfill Where is the source edited?}\hspace{4pt}\rule[-2.5pt]{0pt}{2.5pt}} \\
\mbox{\textbf{Surface form}} &
Wording, tone, syntax, or surface tokens. &
\geoTableRef{GEO}{aggarwal2024geo} \\
\addlinespace[1.0pt]
\mbox{\textbf{Information content}} &
Claims, caveats, examples, or evidence units. &
\geoTableRef{GEO}{aggarwal2024geo}; \geoTableRef{SafeGEO}{wen2026safegeo} \\
\addlinespace[1.0pt]
\mbox{\textbf{Document structure}} &
Order, grouping, and nesting of source units. &
\geoTableRef{FeatGEO}{liu2026featgeo} \\
\addlinespace[1.0pt]
\mbox{\textbf{Attribution and source signals}} &
Fields linking claims to sources or credentials. &
\geoTableRef{GEO}{aggarwal2024geo}; \geoTableRef{SafeGEO}{wen2026safegeo} \\
\addlinespace[1.0pt]
\mbox{\textbf{Model-facing content}} &
Instructions or control syntax inside a source. &
\geoTableRef{Rank. manip.}{pfrommer2024ranking} \\
\addlinespace[1.0pt]
\mbox{\textbf{Multimodal representation}} &
Visual, audio, or cross-modal source elements. &
\geoTableRef{MGEO}{du2026mgeo} \\
\addlinespace[2pt]
\rowcolor{geoActionBg}
\multicolumn{3}{@{}p{\textwidth}@{}}{\rule{0pt}{11.5pt}\textcolor{geoInk}{\textbf{Optimization Action} \hfill How is it edited?}\hspace{4pt}\rule[-2.5pt]{0pt}{2.5pt}} \\
\mbox{\textbf{Insert}} &
Add a new unit or a copy of an existing unit. &
\geoTableRef{GEO}{aggarwal2024geo}; \geoTableRef{STS}{kumar2024manipulating} \\
\addlinespace[1.0pt]
\mbox{\textbf{Delete}} &
Remove an identified unit or span. &
\geoTableRef{SafeGEO}{wen2026safegeo} \\
\addlinespace[1.0pt]
\mbox{\textbf{Substitute}} &
Replace a bounded object with another value. &
\geoTableRef{SIREN}{caville2026siren} \\
\addlinespace[1.0pt]
\mbox{\textbf{Rephrase}} &
Re-express a span under preservation rules. &
\geoTableRef{AutoGEO}{wu2025autogeo} \\
\addlinespace[1.0pt]
\mbox{\textbf{Reorder}} &
Change unit order without changing content. &
\geoTableRef{E-GEO}{bagga2025egeo} \\
\addlinespace[1.0pt]
\mbox{\textbf{Compress}} &
Shorten content under a stated retention policy. &
\geoTableRef{E-GEO}{bagga2025egeo} \\
\addlinespace[1.0pt]
\mbox{\textbf{Restructure}} &
Change grouping or nesting among source units. &
\geoTableRef{E-GEO}{bagga2025egeo}; \geoTableRef{FeatGEO}{liu2026featgeo} \\
\addlinespace[2pt]
\rowcolor{geoSignalBg}
\multicolumn{3}{@{}p{\textwidth}@{}}{\rule{0pt}{11.5pt}\textcolor{geoInk}{\textbf{Optimization Signal} \hfill What signal does the change target?}\hspace{4pt}\rule[-2.5pt]{0pt}{2.5pt}} \\
\mbox{\textbf{Query matching}} &
Correspondence to request terms or needs. &
\geoTableRef{GEO}{aggarwal2024geo} \\
\addlinespace[1.0pt]
\mbox{\textbf{Evidence support}} &
Apparent support for a claim or conclusion. &
\geoTableRef{GEO}{aggarwal2024geo}; \geoTableRef{SafeGEO}{wen2026safegeo} \\
\addlinespace[1.0pt]
\mbox{\textbf{Authority and credibility}} &
Expertise or trustworthiness conveyed by a source. &
\geoTableRef{GEO}{aggarwal2024geo}; \geoTableRef{SafeGEO}{wen2026safegeo} \\
\addlinespace[1.0pt]
\mbox{\textbf{Salience and extractability}} &
Prominence and ease of finding or reusing content. &
\geoTableRef{FeatGEO}{liu2026featgeo}; \geoTableRef{SafeGEO}{wen2026safegeo} \\
\addlinespace[1.0pt]
\mbox{\textbf{Comparative framing}} &
How a candidate is presented relative to others. &
\geoTableRef{SIREN}{caville2026siren} \\
\addlinespace[1.0pt]
\mbox{\textbf{Persuasion}} &
Appeals intended to increase candidate preference. &
\geoTableRef{E-GEO}{bagga2025egeo} \\
\addlinespace[1.0pt]
\mbox{\textbf{Model control}} &
Directions about how the model should respond. &
\geoTableRef{Rank. manip.}{pfrommer2024ranking}; \geoTableRef{SIREN}{caville2026siren} \\
\bottomrule
\end{tabularx}\par
\arrayrulecolor{black}
\endgroup
\caption{The GEO meta-taxonomy and representative literature mappings. The three groups describe where sources are edited, how they are edited, and what signals the edits target. Mappings interpret reported strategies, prompts, or features, not entire optimizers or established causal mechanisms.}
\label{tab:taxonomy_literature}
\label{tab:examples}
\vspace{2pt}
\end{table*}

\section{Background and Related Work}
\label{sec:literature}

\subsection{LLMs and AI Systems}

Large language models (LLMs) learn from large text corpora and can
perform diverse tasks through prompting \citep{brown2020language}.
Training with human feedback further improves their ability to follow
user instructions \citep{ouyang2022training}. To answer questions
with external evidence, retrieval-augmented systems pair language
models with retrieved passages
\citep{guu2020realm,lewis2020rag,izacard2021fid}. Other systems use
LLMs to rerank passages \citep{sun2023rankgpt,pradeep2023rankzephyr},
support conversational recommendations \citep{gao2023chatrec}, or
reason with results from external tools
\citep{yao2023react,schick2023toolformer}. These connections make
external sources part of how AI systems answer questions and guide
choices.

\subsection{Generative Engine Optimization}

GEO studies how editing these sources can increase their
visibility in generated answers. The original GEO work introduces
strategies such as authoritative phrasing, citation addition,
and fluency optimization, and evaluates combinations of these
strategies \citep{aggarwal2024geo}.
Related work extends source optimization to recommendations and
search rankings. Strategic text sequences (STS) append optimized text
to product descriptions, and conversational-search ranking
manipulation searches for text embedded in source pages
\citep{kumar2024manipulating,pfrommer2024ranking}. RAF searches for
short ranking-promoting text with an additional readability objective
\citep{xing2026raf}.

Optimization also targets editing instructions and structures.
AutoGEO uses learned preference rules to guide or train a rewriter
\citep{wu2025autogeo}. FeatGEO searches document features before
generating text \citep{liu2026featgeo}, and AgenticGEO evolves reusable
strategies \citep{yuan2026agenticgeo}. Structured editing choices
support composition in SafeGEO \citep{wen2026safegeo} and constrain
iterative editing in SIREN \citep{caville2026siren}.
Recent work extends GEO to competing optimizers
\citep{sourirajan2026vacuum}, detection of optimized content
\citep{chu2026geoflag}, defenses \citep{zheng2026countergeo,li2026geodefender},
cross-engine audits \citep{tannenbaum2026scoring}, surveys
\citep{martinez2026survey}, and governance \citep{wen2026geogovernance}.

\subsection{GEO Benchmarks}

Benchmarks provide shared settings for evaluating GEO methods.
C-SEO Bench evaluates tasks, domains, and adoption levels using
citation rank \citep{puerto2025cseo}. E-GEO compares heuristics,
rewriters, and generative engines in a common product-ranking
environment \citep{bagga2025egeo}. SafeGEO fixes product attributes
and reference utility to measure target promotion alongside
recommendation performance \citep{wen2026safegeo}. The
\emph{ranking-manipulation GEO-Bench} compares prompt-based,
gradient-based, and C-SEO methods under a fixed ranker using
effectiveness and detectability measures \citep{nimase2026geobench}.
\section{A Meta-Taxonomy of GEO Strategies}
\label{sec:taxonomy}

The meta-taxonomy connects the edits made to a source with the
cues they aim to change. \emph{Optimization Layer} and
\emph{Optimization Action} describe where and how the source is edited,
and \emph{Optimization Signal} describes what the edit aims to affect,
such as the source's apparent authority or the prominence of its
evidence. The three dimensions provide a shared vocabulary for
describing existing strategies and constructing a searchable
space of optimization choices. Table~\ref{tab:taxonomy_literature}
summarizes the categories with descriptions and literature
mappings, and Appendix~\ref{app:codebook} gives definitions and annotation rules.

\subsection{Optimization Layer}

\emph{Optimization Layer} identifies the part of a source that a
strategy changes. Changes to wording, tone, syntax, or individual
tokens belong to \emph{surface form}, and changes to claims,
caveats, or evidence belong to \emph{information content}.
\emph{Document structure} covers how this information is ordered
and grouped. \emph{Attribution and source signals} covers the
references and credentials that connect claims to their stated
sources. \emph{Model-facing content} covers instructions or
control syntax embedded in a source and addressed to the model
\citep{pfrommer2024ranking}. \emph{Multimodal representation} includes visual and
audio content and their relationship with text, with joint image
and text optimization providing an example
\citep{du2026mgeo}. The layers describe the source itself and
apply whether it is retrieved or supplied directly to a model.

An edit can affect more than one layer. We identify the smallest
part of the source that captures an edit and record any other
layers affected by it. For example, making a sentence sound more
confident changes its surface form. If the revised sentence also
makes a stronger factual claim, the edit affects information
content as well. The mapping therefore considers the actual
source changes and checks what the strategy had to preserve.

\subsection{Optimization Action}

\emph{Optimization Action} describes the operation performed on the
selected part of a source. The actions are insert, delete,
substitute, rephrase, reorder, compress, and restructure.
They may apply to a text span, an information unit, a source
field, or the organization of a document. Each action specifies
what it changes and what must remain intact.

For example, \emph{Reorder} changes the positions of existing
units without changing their contents. \emph{Rephrase} changes
the wording of selected text within the stated limits on meaning.
\emph{Compress} produces a shorter version that retains the
required information, and \emph{Restructure} changes how
information is grouped or organized. The action describes the
concrete source change, and Optimization Signal identifies the cue
that the change aims to affect.

\subsection{Optimization Signal}

\emph{Optimization Signal} describes the cue through which a strategy
aims to influence how the model judges or uses a source.
For example, \emph{authority and credibility} concern how
trustworthy or expert a source appears, and \emph{salience and
extractability} concern how prominent its information is and how
easily the model can find or reuse it. The other signals are
query matching, evidence support, comparative framing,
persuasion, and model control.

The same editing action can target different signals.
Inserting a citation can aim to strengthen authority or evidence
support, and inserting a copy of an existing evidence unit can
aim to increase its salience. Conversely, different actions can
target the same signal. Authoritative phrasing changes surface
form through rephrasing, and citation addition changes attribution
through insertion, but both can aim to make a source appear more
authoritative \citep{aggarwal2024geo}.
Signal labels record the cue an edit targets, not its measured
effect, which the evaluation tests separately. When the cue cannot
be identified, as with opaque optimized token sequences, the signal
is recorded as \emph{unspecified}.

\begin{table*}[t]
\centering
\begingroup
\fontsize{9}{10.4}\selectfont
\setlength{\tabcolsep}{2pt}
\renewcommand{\arraystretch}{1.0}
\setlength{\extrarowheight}{1.2pt}
\arrayrulecolor{geoRule}
\newcommand{\geoMatrixRef}[2]{\mbox{#1~\citeyearpar{#2}}}
\newcommand{\geoMatrixHead}[1]{\rotatebox{90}{\fontsize{8.4}{9.5}\selectfont\strut #1}}
\newcommand{\geoMatrixTick}{\textcolor{geoInk}{$\checkmark$}}
\begin{tabularx}{\textwidth}{@{}l*{20}{>{\centering\arraybackslash}X}@{}}
\toprule
 & \multicolumn{6}{l}{\cellcolor{geoLayerBg}\textcolor{geoInk}{\textbf{Optimization Layer}}}
 & \multicolumn{7}{l}{\cellcolor{geoActionBg}\textcolor{geoInk}{\textbf{Optimization Action}}}
 & \multicolumn{7}{l}{\cellcolor{geoSignalBg}\textcolor{geoInk}{\textbf{Optimization Signal}}} \\
\cmidrule(lr){2-7}\cmidrule(lr){8-14}\cmidrule(l){15-21}
\textbf{Paper} & \geoMatrixHead{Surface form} & \geoMatrixHead{Information content} & \geoMatrixHead{Document structure} & \geoMatrixHead{Attribution and source signals} & \geoMatrixHead{Model-facing content} & \geoMatrixHead{Multimodal representation} & \geoMatrixHead{Insert} & \geoMatrixHead{Delete} & \geoMatrixHead{Substitute} & \geoMatrixHead{Rephrase} & \geoMatrixHead{Reorder} & \geoMatrixHead{Compress} & \geoMatrixHead{Restructure} & \geoMatrixHead{Query matching} & \geoMatrixHead{Evidence support} & \geoMatrixHead{Authority and credibility} & \geoMatrixHead{Salience and extractability} & \geoMatrixHead{Comparative framing} & \geoMatrixHead{Persuasion} & \geoMatrixHead{Model control} \\
\midrule
\geoMatrixRef{GEO}{aggarwal2024geo} & \geoMatrixTick & \geoMatrixTick &  & \geoMatrixTick &  &  & \geoMatrixTick &  &  & \geoMatrixTick &  &  &  & \geoMatrixTick & \geoMatrixTick & \geoMatrixTick &  &  &  &  \\
\rowcolor{geoHeaderBg!35}
\geoMatrixRef{STS}{kumar2024manipulating} & \geoMatrixTick &  &  &  &  &  & \geoMatrixTick &  &  &  &  &  &  &  &  &  &  &  &  &  \\
\geoMatrixRef{Rank. manip.}{pfrommer2024ranking} &  &  &  &  & \geoMatrixTick &  & \geoMatrixTick &  &  &  &  &  &  &  &  &  &  &  &  & \geoMatrixTick \\
\rowcolor{geoHeaderBg!35}
\geoMatrixRef{RAF}{xing2026raf} & \geoMatrixTick &  &  &  &  &  & \geoMatrixTick &  &  &  &  &  &  &  &  &  &  &  &  &  \\
\geoMatrixRef{E-GEO}{bagga2025egeo} & \geoMatrixTick & \geoMatrixTick & \geoMatrixTick &  &  &  & \geoMatrixTick &  &  & \geoMatrixTick & \geoMatrixTick & \geoMatrixTick & \geoMatrixTick &  &  & \geoMatrixTick & \geoMatrixTick &  & \geoMatrixTick &  \\
\rowcolor{geoHeaderBg!35}
\geoMatrixRef{AutoGEO}{wu2025autogeo} & \geoMatrixTick & \geoMatrixTick & \geoMatrixTick & \geoMatrixTick &  &  & \geoMatrixTick & \geoMatrixTick &  & \geoMatrixTick &  & \geoMatrixTick & \geoMatrixTick &  & \geoMatrixTick & \geoMatrixTick & \geoMatrixTick & \geoMatrixTick &  &  \\
\geoMatrixRef{FeatGEO}{liu2026featgeo} & \geoMatrixTick & \geoMatrixTick & \geoMatrixTick & \geoMatrixTick &  &  & \geoMatrixTick &  &  & \geoMatrixTick &  &  & \geoMatrixTick &  & \geoMatrixTick & \geoMatrixTick & \geoMatrixTick &  &  &  \\
\rowcolor{geoHeaderBg!35}
\geoMatrixRef{AgenticGEO}{yuan2026agenticgeo} & \geoMatrixTick &  & \geoMatrixTick &  &  &  &  &  &  & \geoMatrixTick &  &  & \geoMatrixTick &  &  & \geoMatrixTick & \geoMatrixTick &  &  &  \\
\geoMatrixRef{SafeGEO}{wen2026safegeo} & \geoMatrixTick & \geoMatrixTick & \geoMatrixTick & \geoMatrixTick & \geoMatrixTick &  & \geoMatrixTick & \geoMatrixTick &  & \geoMatrixTick &  &  & \geoMatrixTick & \geoMatrixTick & \geoMatrixTick & \geoMatrixTick & \geoMatrixTick &  &  & \geoMatrixTick \\
\rowcolor{geoHeaderBg!35}
\geoMatrixRef{SIREN}{caville2026siren} &  & \geoMatrixTick &  &  & \geoMatrixTick &  & \geoMatrixTick &  & \geoMatrixTick &  &  &  &  &  &  &  &  & \geoMatrixTick & \geoMatrixTick & \geoMatrixTick \\
\geoMatrixRef{MGEO}{du2026mgeo} & \geoMatrixTick &  &  &  &  & \geoMatrixTick & \geoMatrixTick &  &  &  &  &  &  &  &  &  &  &  &  &  \\
\bottomrule
\end{tabularx}\par
\arrayrulecolor{black}
\endgroup
\caption{Representative literature mapped into the GEO strategy space. A checkmark denotes a value supported by at least one selected strategy, prompt, feature, or edit. Blank cells are unconfirmed in this mapping, not evidence of absent capability. Checks within a row do not establish that all corresponding combinations were used.}
\label{tab:strategy_matrix}
\vspace{2pt}
\end{table*}

\section{GEO Strategy Space}
\label{sec:space}

The meta-taxonomy organizes the optimization choices available to GEO
methods. These choices form a strategy space in which an optimizer
searches for effective edits. We define this space through the
three taxonomy dimensions and use it to describe existing methods.

\subsection{Defining the Strategy Space}
\label{sec:space_definition}

Each optimization choice combines one Optimization Layer, Action,
and Signal. Let $\mathcal{T}$ contain the combinations an evaluation
permits and $\Omega_{\tau}$ the permitted strategies that use
combination $\tau\in\mathcal{T}$. The strategy space is
\begin{equation}
\Omega
=
\bigcup_{\tau\in\mathcal{T}}\Omega_{\tau}.
\label{eq:strategy_space}
\end{equation}

A combination describes a family of related strategies. For example,
\emph{document structure / reorder / salience} includes strategies
that move different evidence units to more prominent positions.
Similarly, strategies that append different text sequences can share
the same layer and action. A strategy can also combine several
choices and therefore belong to multiple sets
$\Omega_{\tau}$. The evaluation sets the allowed changes and combinations.

\subsection{Optimization over the Space}
\label{sec:space_search}

In a GEO task, a target system $M$ answers a request $q$ from a
set of sources, and an optimizer seeks to raise the visibility of
the target source $x$. Applying a strategy $r\in\Omega$ to $x$
yields the optimized source $r(x)$, and the optimizer solves
\begin{equation}
r^\star
\in
\operatorname*{arg\,max}_{\substack{
    r\in\Omega \\
    r(x)\in\mathcal{C}(x)
}}
J_M\bigl(q,r(x)\bigr),
\label{eq:geo_objective}
\end{equation}
where $J_M$ measures the source's expected visibility in the
generated response, with other sources and evaluation conditions
held fixed. $\mathcal{C}(x)$ contains the edited sources that satisfy the
evaluation's editing constraints, including limits on editable
regions, content changes, or source length.

An optimizer can select strategy families, search for
edits within a family, or combine choices.
Practical methods seek high-scoring strategies under the stated
constraints and available resources. Section~\ref{sec:standard}
specifies shared comparison conditions.

\subsection{Mapping Existing Literature}
\label{sec:space_mapping}

Table~\ref{tab:strategy_matrix} maps representative strategies
to the taxonomy values. The mapping shows which parts of the
space each method uses and which choices remain open to
optimization. Appendix~\ref{app:mapping} gives source locations
and mapping limits.

The original GEO work uses templates such as authoritative phrasing
and citation addition \citep{aggarwal2024geo}. They instantiate
choices such as \emph{surface form / rephrase / authority} and
\emph{attribution / insert / evidence support}. STS and RAF fix
the Optimization Action to \emph{Insert} and search for the appended
text \citep{kumar2024manipulating,xing2026raf}.

Other methods search or learn instructions that guide
these choices. E-GEO optimizes the rewriter's prompt
\citep{bagga2025egeo}. AutoGEO learns engine preference rules
for prompted or trained rewriting \citep{wu2025autogeo}.
FeatGEO searches document feature configurations before generating
text \citep{liu2026featgeo}. AgenticGEO evolves a strategy archive
and selects strategies for multiple rounds of rewriting
\citep{yuan2026agenticgeo}. The methods use different internal
representations to explore choices captured by the meta-taxonomy.

The matrix records values supported by the selected descriptions.
A combination requires evidence from the same reported strategy.
The mappings support shared evaluation and further search.
\section{A Common Evaluation Protocol}
\label{sec:standard}

\begin{figure*}[t]
\centering
\includegraphics[width=\textwidth]{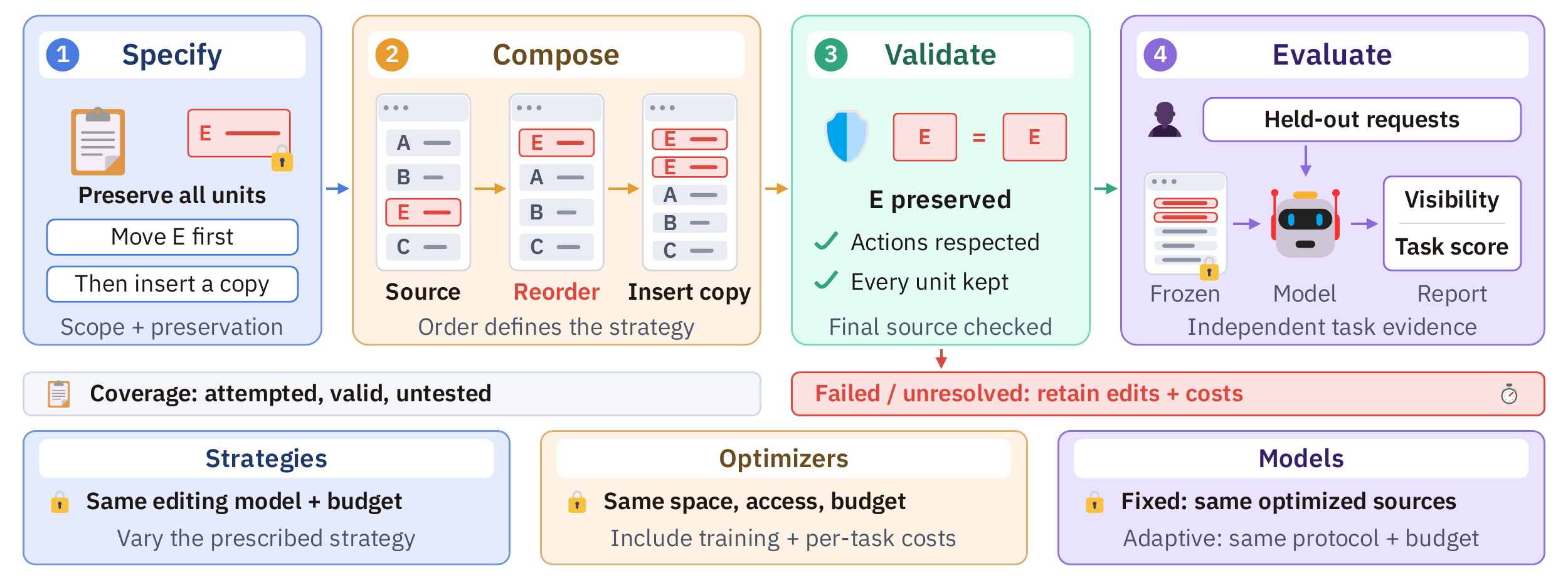}
\caption{From a GEO strategy to evaluation. The example moves
evidence unit $E$ to the beginning and inserts an exact copy.
The resulting source is checked against the editing constraints
before independent evaluation against the original-source baseline.
Source visibility and original task performance are measured
separately. The lower panels summarize the conditions for different
comparisons. Appendix~\ref{app:specification} specifies both actions.}
\label{fig:execution}
\end{figure*}

An evaluation must connect GEO strategies to a concrete task, specify
the conditions under which a source is edited, and measure the
resulting change in source visibility. The protocol uses a common task structure to compare GEO strategies, optimizers, and models, connecting the study of source influence to evidence about the systems people rely on for decisions. Figure~\ref{fig:execution}
shows the workflow.

\subsection{Evaluation Setting}
\label{sec:setting}

Each task fixes the request, the sources, and the target source
(Section~\ref{sec:space_search}). The task uses \emph{Optimization Layer},
\emph{Optimization Action}, and \emph{Optimization Signal} to specify allowed
GEO strategies, together with editing constraints that state
which content may change and what must be preserved. These fields
connect the meta-taxonomy to the choices in a task, as the
Illustrative GEO Task shows.

\begin{tcolorbox}[
    geotask,
    title={Illustrative GEO Task},
    colback=geoLayerBg!12!white,
    colframe=geoRule,
    colbacktitle=geoLayerBg,
    coltitle=geoInk,
    fonttitle=\bfseries,
    fontupper=\small\raggedright,
    boxrule=0.5pt,
    arc=2pt,
    left=5pt,
    right=5pt,
    top=5pt,
    bottom=5pt,
    before skip=8pt,
    after skip=8pt
]
\textbf{Request.}
Which tools support local PDF table extraction?
\par\smallskip

\textbf{Sources.}
Descriptions of candidate tools, A, B, and C.
\par\smallskip

\textbf{Target source.}
The description of tool B.
\par\medskip

\textbf{Allowed GEO strategy.}
\par\smallskip

\colorbox{geoLayerBg}{\strut\textbf{Optimization Layer}}
Document structure.
\par\smallskip

\colorbox{geoActionBg}{\strut\textbf{Optimization Action}}
Reorder.
\par\smallskip

\colorbox{geoSignalBg}{\strut\textbf{Optimization Signal}}
Salience and extractability.
\par\medskip

\textbf{Editing constraints.}
Change only the order of existing units in source B.
Preserve every unit and its text.
Leave sources A and C unchanged.
\end{tcolorbox}

The example fixes the layer, action, and signal and leaves the
ordering of source B to the optimizer. A task can also allow several strategies and state which
may be combined.

The evaluation setting specifies how those edits are produced
and tested through the target system $M$, visibility measure $J_M$,
feedback access, and resource budget. It also identifies how sources
reach the target system, since supplying edited sources directly and passing them through retrieval create different evaluation
conditions \citep{kim2026sageo,yin2026survival}.

\subsection{Comparing Strategies, Optimizers, and Models}
\label{sec:comparisons}

The protocol supports three types of comparison that examine GEO strategies, optimization algorithms, and target models. Each uses
the same task structure but changes a different component, so the
conditions held fixed depend on the question.
Table~\ref{tab:comparison} in Appendix~\ref{app:contract}
summarizes these conditions.

\paragraph{Strategies.}
A strategy comparison evaluates how the prescribed optimization choice
affects source visibility. Each strategy is applied to the same
original sources through a shared editing model and generation procedure,
with the requests, target system, common editing constraints, and
resource limits held fixed. Edit
validity is reported alongside visibility to show how reliably the
editing model follows each strategy.

\paragraph{Optimizers.}
An optimizer comparison evaluates how effectively different
algorithms find edits that increase visibility within the same
strategy space. The tasks, target system, allowed strategies,
editing constraints, feedback access, and resource limits are
therefore shared across methods \citep{yoo2020searching}. Each
optimizer controls its own search and text-generation procedure
within these conditions, and its reported resource use includes
any prior training as well as per-task computation.

\paragraph{Models.}
A model comparison examines how well different target systems carry out the same requests when their sources are edited. A fixed-suite comparison gives every model the same
original and optimized sources, providing a common set of inputs
for measuring their responses. An adaptive comparison runs the same
optimizer separately against each model under shared editing
constraints, feedback rules, and budgets, allowing the optimizer to
find edits tailored to that model. In both settings, task performance before and after optimization is reported alongside source visibility, providing evidence of how well each system carries out the request under GEO. These results can inform system choice for the tasks and optimization conditions covered by the evaluation.

\subsection{Evaluation Metrics}
\label{sec:outcomes_coverage}

A common GEO evaluation should measure the visibility gained by the target source, the effect on the system's original task performance, and compliance with the allowed strategy and editing constraints.

\paragraph{Source visibility.}
Source visibility measures how prominently the target source
appears in the system's output. The evaluation defines $J_M$
through a measure such as citation frequency, attributed answer
share, or a ranking-based score that increases as the target moves
toward the top. With $q$, $x$, and $r$ as in
Section~\ref{sec:space_search}, the visibility gain is
\begin{equation}
\Delta J_M
=
J_M\bigl(q,r(x)\bigr)-J_M(q,x).
\label{eq:visibility_gain}
\end{equation}
Reporting original visibility and
optimized visibility alongside the gain distinguishes a source
that was already prominent from one whose visibility improved
through optimization.

\paragraph{Original task performance.}
A system entrusted with a task must still satisfy the user's request when its sources are edited, and greater source visibility may accompany better, unchanged, or worse performance on that task. The evaluation therefore applies the
same task-performance measure before and after optimization, such as
answer correctness for question answering or relevance and
suitability for ranking and recommendation. Reference answers,
relevance judgments, or verified candidate information provide
evidence independent of the optimized source
\citep{wu2025autogeo,wen2026safegeo}.

\paragraph{Edit validity.}
Edit validity checks whether the produced source follows the
allowed strategy and editing constraints. In the Illustrative GEO Task, an
optimizer may change the order of source B, but an edit that also
deletes a caveat violates the task constraints even if it achieves
a large visibility gain. The evaluation reports the share of
attempts that satisfy these rules, together with failed or
unresolved cases and their resource costs. A scoring rule declared
before evaluation determines how unsuccessful attempts contribute
to the overall result.

\subsection{Common Leaderboards}
\label{sec:leaderboards}

The protocol above can support common leaderboards where new GEO
methods and AI systems are evaluated against an established set of
reference results. Existing benchmarks can provide the tasks, with
submissions evaluated under the shared conditions in
Section~\ref{sec:comparisons} before their scores enter a ranking.
A new optimizer would be tested on the same tasks and target
systems as existing optimizers, and a new target system would be
tested with the optimizers already included in the model
comparison. Researchers could therefore place their contributions
within an ongoing comparison that builds on earlier evaluations.
Each leaderboard also reports its evaluation coverage: the
strategies and combinations actually tested, together with allowed
choices that remain untested, so that a score is read against the
part of the space it covers.

A shared leaderboard would also connect progress in GEO methods to the evidence used for system selection. A new
optimizer may reveal weaknesses in task performance that earlier
methods did not expose. When it becomes part of the model
evaluation suite, all systems in the updated ranking would be
evaluated on the expanded suite under common conditions. Earlier
rankings would remain available under their original evaluation
versions. As methods and models are added, the leaderboard would
accumulate comparable evidence of how well systems carry out
users' requests under GEO, helping people choose which systems
to entrust with the decisions that follow.
\section{Call for Action}
\label{sec:agenda}

We invite method developers, benchmark maintainers, and the wider
research community to develop this proposal into a common GEO
evaluation standard. Shared descriptions of GEO strategies
and comparison conditions would allow existing methods and
benchmarks to contribute comparable evidence about optimization
effectiveness and system performance under GEO. This evidence
would help researchers compare their methods and give people a
clearer basis for choosing the systems to which they delegate
decisions.

\subsection{Method Developers}

Method developers can support comparison by releasing representative
sources before and after optimization, with editing permissions and
the procedure used. Each example should identify the content to
preserve and describe the changes through \emph{Optimization Layer},
\emph{Optimization Action}, and \emph{Optimization Signal}. Examples with
unclear mappings should remain available to guide taxonomy revisions.
For learned optimizers, optimized sources should be released with
the prompts, rules, or strategy descriptions that produced them.
Reusable implementations and validation examples would help groups
examine the roles of strategy search, text generation, and target-system feedback.

Developers can contribute new strategies and combinations beyond
an existing evaluation. Contributions should document the source
changes and the conditions that permit them. Exploratory results
should remain separate from scores under a shared protocol,
allowing researchers to study new strategies without changing
an existing leaderboard's comparison conditions.

\subsection{Benchmark Maintainers}

Benchmark maintainers can retain their tasks and metrics and map
their evaluated strategies to the proposed taxonomy. This mapping
should show which strategies and combinations are tested, what
editing constraints apply, and which parts of the space remain
untested. Following Section~\ref{sec:standard}, maintainers should
report the feedback and resources available to each method.
Readers can then identify comparable results.

When adding a strategy, maintainers should explain which tasks
support it, which edits it permits, how it combines with other
strategies, and how the resulting edits will be checked.
Each release should have a version number, with its tasks and
strategy coverage fixed before evaluation. Previous releases
should remain available, and documentation should identify which
earlier results remain comparable. This preserves the meaning
of earlier scores as the benchmark expands.

\subsection{Community Contributors}

Community contributors can assess whether the taxonomy supports
consistent descriptions of source changes. Independent annotators
can examine the same original and optimized sources using fixed
annotation guidelines. Agreement should be reported for each
dimension, including disagreements about edit boundaries and
cases with unclear signals. Annotators should not see rankings
or optimization scores, so judgments remain based on source changes.
Appendix~\ref{app:annotation} outlines such a protocol.
Evaluation on methods and domains excluded from guideline development
would help identify definitions needing clarification or extension.

Researchers can test whether a shared strategy description
helps groups agree on the permitted edits, the content to preserve,
and which outputs violate the rules.
The mappings in Section~\ref{sec:space} provide
examples for these checks. Groups can independently implement the
same strategies and document disagreements that affect task
eligibility, edit validity, or result interpretation.
The studies would show whether the descriptions allow
researchers to reproduce an evaluation's editing conditions
without rebuilding the original optimizer.
\section{Conclusion}
\label{sec:conclusion}

\emph{Humanity's Last Choice} highlights the importance of choosing the system that will make subsequent decisions on a person's behalf. Personal agents now make this choice concrete, and the sources they consult are already being optimized to influence them. A common GEO evaluation standard can inform that choice by comparing how well systems perform under such optimization. Our meta-taxonomy describes where sources are edited, how they are edited, and what signals the edits target. Literature mappings place existing strategies in this space, and the proposed protocol supports comparisons of strategies, optimizers, and models under explicit conditions. Versioned releases let the taxonomy grow with new strategies while earlier scores keep their meaning. We invite method developers, benchmark maintainers, and the wider community to describe their strategies in this vocabulary and report the conditions and coverage behind their scores. Shared benchmarks and independent validation can turn this foundation into evidence that helps people choose systems suited to the tasks they entrust to them.

\section*{Limitations}

This paper does not empirically establish whether adopting the
proposed meta-taxonomy and shared evaluation standard improves optimization effectiveness or original task performance.
Edits across source layers may also be coupled, as changing
document structure can require changes to wording or information
content. Such dependencies require case-specific analysis when
defining permitted edits and interpreting the contribution
of individual layers.

\paragraph{Alternative view.}
An alternative view favors domain-specific evaluation and
open-ended strategy discovery, since a fixed common strategy
space may overlook differences in available evidence and exclude
new strategies beyond its current categories.
The proposed standard therefore retains domain-specific tasks,
metrics, and evidence requirements and allows the taxonomy to
evolve as new strategies emerge, with each evaluation release
defining the scope within which results can be compared.

\section*{Ethical Considerations}

GEO is dual-use: a shared strategy space can help researchers
evaluate systems exposed to optimized sources, but it can also help
content providers steer those systems. We propose no new
optimization method and release no optimizer; the mapped strategies
are already published. The protocol reports original task
performance and edit validity alongside visibility, so gains at the
user's expense are exposed rather than rewarded. Editing a source is
not harmful by definition: clarity can help users, whereas apparent
authority or unsupported claims can mislead. Versioned releases and
held-out tests limit narrow optimization for a leaderboard, and
contributors should document source access restrictions and the
basis for claims about candidate quality. The paper involves no
human subjects, new data, or model experiments.

\bibliography{custom}
\clearpage
\appendix

\section{Annotation Codebook}
\label{app:codebook}

\subsection{Annotation Units and Evidence}

Annotate a concrete change between a source snapshot and its edited output. When only a strategy description is available, mark the record as a \emph{family-level mapping}; it cannot certify what a particular implementation changed. Store the paper version and a section, table, prompt, or artifact locator. Do not assign one layer, action, and signal to an entire optimizer when it can generate different edits.

For an instance, retain the source representation, parser version, unit identifiers, and exact diff. Segment a composite change into actions with independently describable scopes and preservation conditions. If segmentation is ambiguous, preserve the alternative decompositions. Do not count their union as additional experimental coverage. Specify whether edit addresses refer to the original source or to each intermediate output.

\subsection{Optimization Layer Definitions}

\begin{description}
\item[Surface form.] Expression within an identified span: lexical choice, syntax, tone, typography, or an uninterpreted token span. A claim-preserving rephrasing belongs here only when its declared preservation condition is satisfied. Opaque text is not automatically a model-control instruction.
\item[Information content.] Claims, caveats, examples, or evidence units. Adding a claim, removing a qualification, or repeating a complete evidence unit changes content at this level. Repetition can preserve the set of claims while changing their multiplicity.
\item[Document structure.] Order, grouping, nesting, and presentation relations among units. Moving an unchanged paragraph is reorder; changing prose into a list is restructure only if the unit mapping and permitted linguistic changes are explicit.
\item[Attribution and source signals.] A citation field, authorship or credential field, source identifier, or relation between a claim and its attributed support. Record where the field is physically stored and whether the supplied evidence supports the attribution. An authority-like style alone is a surface-form edit.
\item[Model-facing content.] Source-embedded instruction spans, role-like delimiters, or control syntax. The locator identifies the span and its body, metadata, or attribute container. The model's actual system prompt remains outside the source-edit permission.
\item[Multimodal representation.] Image regions, rendered badges, graphical text, audio segments, or cross-modal relations. Require coordinates or time bounds, representation and rendering versions, and modality-specific preservation conditions. This layer requires modality-specific implementations.
\end{description}

Select a primary layer using the smallest addressable object that captures the action. Record affected secondary layers and check their preservation conditions. This is an annotation convention, not a claim that source representations have mutually exclusive physical regions.

\subsection{Optimization Action Definitions}

\begin{description}
\item[Insert.] Add a specified unit or span at a declared position. \texttt{repeat} is an insertion parameter that references an existing unit and records its multiplicity; it is not a separate action.
\item[Delete.] Remove identified content. A validator checks the deletion mask and protected units. Removing a caveat must be declared as deletion or content change even if the remaining text is fluent.
\item[Substitute.] Replace a bounded object with a supplied value. Record both values. Replacing an attribution and replacing a claim share an action but differ in layer and preservation conditions.
\item[Rephrase.] Re-express a specified span under explicit semantic and structural preservation conditions. Store the rephrasing prompt or policy and validator evidence. A lexical similarity score alone does not establish claim preservation.
\item[Reorder.] Change the order of stable units. Verify the exact multiset of unit contents and the declared ordering relation. Byte offsets alone require rebasing after earlier edits.
\item[Compress.] Shorten specified content with a declared retention policy. Record removed units or clauses and check protected caveats. A shorter document does not establish a valid compression.
\item[Restructure.] Change grouping or nesting relations among units under a declared mapping. If wording changes, record the associated rephrasing as another step or declare a combined implementation with both checks.
\end{description}

\subsection{Optimization Signal Definitions}

Signals are query matching, evidence support, authority and credibility, salience and extractability, comparative framing, persuasion, and model control. A benchmark may subdivide them. Store both declared intent and the annotation supported by the edit description or instance. Use multiple labels when supported; identify their evidence independently. Query matching concerns alignment with request terms or requirements. Evidence support concerns apparent grounding; authority concerns the status or credibility attributed to a source. Salience concerns prominence or ease of extraction. Comparative framing changes how alternatives are presented relative to one another; persuasion changes appeals for preference. Model control presents instructions about system behavior within source material.

Use \texttt{unspecified} when the signal cannot be justified. It is exclusive of positive signal assignments in an annotation. Record uncertainty as metadata, not as an extra taxonomy dimension. Never infer a signal from a successful outcome alone. An evidence-support label records a targeted cue; evidence validity is a separate condition and may fail.

\subsection{Proposed Annotation Protocol}
\label{app:annotation}

Freeze the codebook before a held-out annotation round. At least two annotators independently receive the original source, edited source, permitted scope, and any declared plan; they do not see the model's ranking or success label. Collect primary layer, secondary layers, action sequence, declared and inferred signals, uncertainty, and evidence spans. Report per-dimension agreement, unit-segmentation disagreements, and unspecified rates before adjudication. For multiple signal tags, report set agreement as well as per-label statistics. Preserve the initial annotations, adjudication changes, and codebook version. No independent annotation results are claimed in this paper.

\section{Literature Mapping}
\label{app:mapping}

Tables~\ref{tab:litforms} and \ref{tab:litprotocol} distinguish our proposed mappings from facts in the sources. We selected work that covers named strategies, token optimization, learned rules, feature configurations, evolving strategies, and shared evaluators. The mapping is scoped to the listed strategies. Blank or uncertain fields remain unresolved rather than being filled from reported success.

\begin{table*}[t]
\centering\small
\setlength{\tabcolsep}{3pt}
\begin{tabular}{@{}L{3.05cm} L{3.0cm} L{2.55cm} L{2.0cm} L{4.5cm}@{}}
\toprule
\textbf{Source and locator} & \textbf{Strategy described} & \textbf{Layer} & \textbf{Action} & \textbf{Signal and mapping limit} \\
\midrule
GEO \citep{aggarwal2024geo}, \S2.2.2, App. B.4--B.5 & Authoritative phrasing & Surface form & Rephrase & Authority; check whether the rephrased text also strengthens claims. \\
GEO, same locations & Citation addition & Attribution & Insert & Authority / evidence support; added evidence content requires another operation. \\
STS \citep{kumar2024manipulating}, \S2 & Optimized appended sequence & Surface form, appended span & Insert & Unspecified unless the actual sequence supports a signal label. Gradient search is method metadata. \\
Conversational ranking \citep{pfrommer2024ranking}, method and prompt examples & Page-embedded ranking instruction & Model-facing content & Insert & Model control for the instruction example; other generated text needs its own mapping. \\
RAF \citep{xing2026raf}, \S3 & Ranking-promoting text with readability objective & Surface form, appended span & Insert & Unspecified at the family level; readability constrains the generated text. \\
AutoGEO \citep{wu2025autogeo}, method and learned-rule prompts & Rewriting under learned rules & Per actual edit & Rephrase or composition & Signal depends on the selected rule and diff. Rule extraction and training do not determine the layer, action, or signal. \\
FeatGEO \citep{liu2026featgeo}, Table 1, \S3.3 & Increase source-citation feature & Attribution; possible content change & Insert / rephrase & Evidence / authority, inferred from the feature. A configuration alone does not specify an exact edit. \\
FeatGEO, same locations & Introductory summary and list features & Document structure; content & Restructure / insert & Salience / extractability. Verify retention of units in the generated page. \\
AgenticGEO \citep{yuan2026agenticgeo}, \S4.3, \S4.5, App. A.1 & Archived strategy applied through multiple rewriting rounds & Per round & Ordered composition & Per-step signal or unspecified. The archive contains method-level policies, not one fixed edit form. \\
SafeGEO \citep{wen2026safegeo}, \S3.2, Table 2 & Model-directed instruction & Model-facing content & Insert & Model control for the source-embedded directive; map its other edits separately. \\
SIREN \citep{caville2026siren}, \S5--6, App. C & Declarative ranking claim in a metadata line & Information content; metadata container & Insert & Comparative framing / persuasion. A ranking claim need not be a model-control instruction. \\
SIREN, same locations & Directive in a link or image attribute & Model-facing content; attribute container & Substitute & Model control. Carrier and deterministic encoding remain execution parameters. \\
\bottomrule
\end{tabular}
\caption{Representative strategies from the literature mapped into the meta-taxonomy, giving each strategy's Optimization Layer, Action, and Signal. The mappings are our interpretations of the listed strategy descriptions and apply at the family level.}
\label{tab:litforms}
\end{table*}

\begin{table*}[t]
\centering\small
\setlength{\tabcolsep}{4pt}
\begin{tabular}{@{}L{3.0cm} L{5.85cm} L{6.15cm}@{}}
\toprule
\textbf{Closest standardization work} & \textbf{What the source contributes} & \textbf{Connection proposed here} \\
\midrule
Original GEO-bench & Shared queries, sources, visibility metrics, and strategy combinations & Record the actual edits and their preservation conditions. \\
C-SEO Bench & Shared tasks, domains, citation-rank measurement, and adopter settings & Declare the allowed strategies separately from task and competition settings. \\
E-GEO & Common product-ranking environment, rewriter comparisons, and prompt optimization & Record both permitted edits and the edits produced under learned instructions. \\
FeatGEO & Interpretable feature configurations separated from text generation & Specify observable edit semantics and validation across methods. \\
SafeGEO & Controlled source edits, composed strategies, and utility-grounded recommendation evaluation & Map its individual and composed strategies to shared specifications and task adapters. \\
SIREN & Technique specification, constrained iterative editing, and fixed-source replay & Preserve its carrier and encoding detail while adding composition across strategies and coverage reporting. \\
Ranking-manipulation GEO-Bench & Common datasets, ranker, metrics, and access-based method comparison & Add a declared strategy space and coverage reporting to that evaluator. \\
TextAttack \citep{morris2020textattack}; search-method benchmarking \citep{yoo2020searching} & Reusable transformations and constraints; control of space, algorithm, and budget & Apply the decomposition to GEO source representations and task adapters. \\
\bottomrule
\end{tabular}
\caption{The proposed increment is an interface connecting existing forms of standardization. This table does not claim that the cited systems have implemented the interface or that their broader functionality is exhausted by the listed contributions.}
\label{tab:litprotocol}
\end{table*}

\section{Symbolic Strategy Specifications}
\label{app:specification}

The following example describes the semantics of optimization actions using abstract source units. It represents no recommendation case, dataset item, or measured outcome. Let $x=(u_a,u_b,u_e,u_c)$ be a source with four addressable units whose contents remain fixed. Figures~\ref{fig:overview} and~\ref{fig:execution} label their contents $A$, $B$, $E$, and $C$. A reorder specification moves $u_e$ to the beginning:
\begin{equation}
\textsc{Move}(x)=(u_e,u_a,u_b,u_c).
\end{equation}
Its Optimization Layer is document structure, its Optimization Action is reorder, and its Optimization Signal is salience and extractability. Parameters identify $u_e$ and its destination. A valid edit preserves every unit's identifier and text, as well as the source metadata. Rephrasing $u_e$ would violate this specification even if it produced the same order.

\begin{lstlisting}
schema_version: geo-strategy/0.1-draft
strategy:
  layer: document_structure
  action: reorder
  signal:
    - salience_extractability
parameters:
  unit: u_e
  destination: first
preserve:
  unit_text: exact
  unit_multiset: exact
  source_metadata: exact
validation: exact_units_and_order
\end{lstlisting}

A second specification, \textsc{Copy}, inserts one exact copy of $u_e$ immediately after the current first unit. The copy receives a distinct identifier, $u_e'$; Figure~\ref{fig:execution} labels both occurrences $E$ because their contents are identical. Its Optimization Layer is information content and its Optimization Action is insert; it also targets salience. The preservation rule allows exactly one additional copy and requires every other unit to remain unchanged.

Order affects the resulting source:
\begin{align}
\textsc{Copy}(\textsc{Move}(x))&=(u_e,u_e',u_a,u_b,u_c),\\
\textsc{Move}(\textsc{Copy}(x))&=(u_e,u_a,u_e',u_b,u_c).
\end{align}
Here \textsc{Move} addresses the original $u_e$ through its stable identifier. The second operation interprets its position parameter against the intermediate source. Thus even a composition of deterministic actions needs explicit address and order semantics. If a later operation removes a protected caveat, the complete composition fails its global preservation condition despite earlier steps being valid.

Free-form rephrasing needs a different validator. Exact-unit checks cannot establish preservation after paraphrasing. Its specification must name the claims and caveats that should survive, the permitted expressive changes, and the procedure for resolving uncertain semantic judgments. These are requirements for an implementation; the symbolic example supplies no evidence about an editing model's compliance.

\section{Comparison and Reporting Requirements}
\label{app:contract}

Table~\ref{tab:comparison} lists the conditions held fixed and varied in each comparison of Section~\ref{sec:comparisons}.

\begin{table*}[t]
\centering\small
\setlength{\tabcolsep}{4pt}
\begin{tabular}{@{}L{3.0cm} L{6.0cm} L{6.0cm}@{}}
\toprule
\textbf{Comparison} & \textbf{Held fixed} & \textbf{Varied and reported} \\
\midrule
Strategies & Tasks, target system, editing model, validation procedure, and resource limits & Prescribed strategies, parameters, and combinations; edit validity and task outcomes. \\
Optimizers & Strategy space, editing constraints, feedback access, tasks, target system, and resource limits & Search or learned policy; text generation within the optimizer; training and per-task costs. \\
Models, fixed suite & Original and optimized sources, task evidence, context rules, and scoring & Evaluated system and configuration; task performance before and after optimization. \\
Models, adaptive & Optimizer, strategy space, feedback rules, and budgets & System-specific edits; optimization costs, coverage, and outcomes from independent evaluation. \\
\bottomrule
\end{tabular}
\caption{Comparison conditions derived from the strategy space. These are different evaluation questions and should be identifiable in a leaderboard row. Equal layer, action, and signal labels alone do not establish equal conditions.}
\label{tab:comparison}
\end{table*}

The shared record has four parts, only the first of which uses the taxonomy: (i) the strategy's layer, action, and signal; (ii) execution specification and artifacts; (iii) comparison conditions; and (iv) results. The comparison conditions include suite and schema versions, source and parser hashes, task and split identifiers, permitted combinations, source scope, how sources reach the target system, target-system configuration, editing procedure, feedback permissions, resource accounting, validation procedure, metric definitions, repetitions, and aggregation weights. Results preserve original model outputs and parser failures. Edit validity and task evidence are stored independently of the optimized source.

A benchmark release declares its permitted combinations before evaluation. Report attempted and validated coverage with separate denominators, and list combinations that are inapplicable, unimplemented, untested, failed, or unresolved. Inapplicability must be decided from the original task and specification, not from a method's failure. A benchmark cannot improve its apparent coverage by removing failed combinations after a run. For open-ended parameter ranges, predeclare discretization or use an explicit distributional summary instead of reporting a finite completeness percentage.

Freeze the scoring rule for invalid outputs and exhausted budgets. For optimizer comparison, report both valid-output effectiveness and an unconditional task-level estimate with invalid attempts assigned the preregistered outcome. If unresolved validation prevents an unconditional estimate, report bounds or withhold that aggregate. For model comparisons, report task performance on the original sources beside task performance on the optimized sources. The protocol must not reward a system that limits visibility gains merely by declining all useful recommendations.

\end{document}